\documentclass[aps,amsmath,amssymb,superscriptaddress,twocolumn,showpacs,12pts, floatfix, showpacs,intlimits]{revtex4-1}

\usepackage[english]{babel}
\usepackage[utf8]{inputenc}

\usepackage{amsthm}
\usepackage{mathtools}
\usepackage{physics}
\usepackage{xcolor}
\usepackage{graphicx}
\usepackage[left=23mm,right=13mm,top=35mm,columnsep=15pt]{geometry} 
\usepackage{adjustbox}
\usepackage{placeins}
\usepackage[T1]{fontenc}
\usepackage{lipsum}
\usepackage{csquotes}

\usepackage{orcidlink}

\begin{document}

\title {Constant sensitivity birefringence metrology using vector vortex beams}

\author{Gabriela Flores-Cova~\orcidlink{0009-0004-6814-0672}}
\affiliation{ICFO – Institut de Ciències Fotòniques, The Barcelona Institute of Science and Technology, 08860 Castelldefels Spain}
\affiliation{Centro de Investigaciones en Óptica, A.C., Loma del Bosque 115, Colonia Lomas del Campestre, 37150 León, Guanajuato, México.}%

\author{Daniel Salamanca- Rold\'an~\orcidlink{0009-0007-5958-162X}}
\affiliation{ICFO – Institut de Ciències Fotòniques, The Barcelona Institute of Science and Technology, 08860 Castelldefels Spain}
\affiliation{Centro de Investigaciones en Óptica, A.C., Loma del Bosque 115, Colonia Lomas del Campestre, 37150 León, Guanajuato, México.}%

\author{ Carmelo Rosales-Guzm\'an~\orcidlink{0000-0002-0321-0877}}
\affiliation{Centro de Investigaciones en Óptica, A.C., Loma del Bosque 115, Colonia Lomas del Campestre, 37150 León, Guanajuato, México.}%

\author{Daniel F. Urrego~\orcidlink{0000-0002-9067-7909}}
\affiliation{ICFO – Institut de Ciències Fotòniques, The Barcelona Institute of Science and Technology, 08860 Castelldefels}%

\author{Juan P. Torres~\orcidlink{0000-0002-4454-6676}}
\affiliation{ICFO – Institut de Ciències Fotòniques, The Barcelona Institute of Science and Technology, 08860 Castelldefels}%
\affiliation{Department of Signal Theory and Communications, Universitat Politecnica de Catalunya, 08034 Barcelona, Spain}%


\begin{abstract} 
Differential Interference Contrast (DIC) microscopy and chiral analysis are two imaging techniques that measure the birefringence, i.e., the phase difference introduced by a sample on two orthogonal polarizations. Conventional approaches employ Gaussian beams and infer birefringence from polarization changes, resulting in phase-estimation sensitivities that depend on the unknown phase. We demonstrate here a new type of birefringence detector. It makes use of a vector vortex beam, a type of structured light endowed with optical modes that carry opposite orbital angular momentum (OAM). Using quantum estimation theory tools, we demonstrate that the sensitivity of phase estimation is independent of the value of the unknown phase, and can be even better, in principle, than the conventional approach. We experimentally validate the proposed scheme, demonstrating the potential of structured light for robust and uniform birefringence sensing.
\end{abstract}
\maketitle
\section{Introduction}
\label{intro}
We consider two imaging schemes that measure the birefringence of a sample. Although the physical origin of the phase difference in each scheme is different, they can be described in a unified manner. The first scheme is Differential Interference Contrast (DIC) microscopy. It estimates the thickness variations of a phase object by transforming local thickness gradients into optical path differences~\cite{murphy2003book,Terborg2016}. In DIC, a specimen is sampled by a pair of closely spaced optical beams with orthogonal polarizations. The two beams acquire different phase shifts since they traverse the sample at different, but close, locations. As a result, after recombination of the two beams with orthogonal polarizations into a single beam, the state of polarization of the output optical beam changes with respect to the input state of polarization. 

The second scheme is Chiral analysis, which detects and quantifies the presence in a solution of specific enantiomers of chiral molecules~\cite{barron2004book}. Chiral molecules show circular birefringence (optical activity), that is the difference between the refractive index for left- and right-circularly polarized light. The resulting phase difference generates a rotation of the plane of polarization of a linearly-polarized optical beam. Most biological molecules, such as proteinogenic amino acids, carbohydrates, nucleosides, antibiotics and vitamins, show circular birefringence. Modern nanoscale fabrication techniques have made it possible to design and manufacture metamaterials with engineered geometric chirality, producing a significantly enhanced chiral response with visible light, well beyond those typically observed in natural systems~\cite{liningerAdvMat2023}.

The {\it standard} procedure to measure birefringence is to measure the {\it global} polarization change of a coherent light beam with a Gaussian spatial shape. The precision of phase difference estimation depends on the specific value of the unknown phase. In order to gain sensitivity, light detection operates close to, but not exactly, at total extinction, i.e., detection of output light with polarization orthogonal to the input polarization. Under these conditions, the precision of phase estimation is (see Section~\ref{metrological_advantage} below) $\text{Var}({\hat{\theta}}) =1/(N_0 T_0)$, where $\theta$ is the phase difference, ${\hat{\theta}}$ is the phase estimator, $\text{Var}$ designates the statistical variance of the phase estimation, $1-T_0$ designate losses and $N_0$ is the number of photons probing the sample. For coherent states, this measurement scheme is optimum, i.e., it is a bound to the best precision that any other measurement scheme could achieve.

To make the sensitivity of birefringence estimation independent of the unknown phase difference $\theta$, we put forward an imaging system that makes use of vector vortex beams. Vector vortex beams are described as a non-separable superposition of the polarization and spatial degrees of freedom, since the state of polarization changes across the wavefront of the beam. This is why this type of optical beams are sometimes described as {\it non-separable} optical beams~\cite{Zela2014,Valles2014,aiello2022,Urrego2020}. We consider here a type of vector vortex beam that is the coherent superposition of two optical beams with orthogonal polarizations and orthogonal spatial shapes. The transverse spatial shapes associated to each polarization read as ${\bf E}=E_0(\rho) \exp(\pm i m \varphi)$ where $(\rho,\varphi$) are cylindrical coordinates, $E_0$ is the radial profile of the spatial shape and $m$ is the topological charge. These optical beams contain $m\hbar$ orbital angular momentum (OAM) per photon~\cite{Yao2011,torner20011book}. 

This type of optical beams have been considered previously for measuring phase differences in interferometric setups, such as a Michelson interferometer~\cite{Emile2017,Verma2019,zhang2022,kerschbaumer2022}. Projection of the vector vortex beam into diagonal or anti-diagonal polarization generates a $2|m|$ petal-like structure of the intensity in the transverse plane. Any phase difference $\theta$ between the two OAM optical beams generates a rotation $\theta/(2|m|)$ of the intensity pattern that can be readily measured. We demonstrate here that this measurement scheme can be used to measure birefringence, by substituting the detection of a polarization change with the measurement of the rotation of an optical beam's intensity pattern. 

In Section~\ref{metrological_advantage} we demonstrate that birefringence metrology based on the use of vector vortex beams can provide a metrological advantage for phase estimation in DIC and Chiral analysis for certain ranges of values of the unknown phase $\theta$. For this, we make use of rigorous parameter estimation theory, a tool that provides a reliable way to analyze the precision and the resolution attainable in the estimation of an unknown parameter ~\cite{kay1993book,vandeBos2007book,Motka2016}. In Section~\ref{experimental_setup} we describe the main features of the experimental setup used. Section~\ref{experimental_results} shows experimental results that demonstrate the feasibility of measuring birefringence with vector vortex beams. We make use of two types of samples: a liquid crystal variable retarder (LCVR) introduces birefringence in a controllable manner and allows us to calibrate our system, and a birefringent resolution target from Thorlabs. In Section~\ref{conlusions}, we provide a brief analysis of the relevance of our work in the general context of imaging and estimation using structured light, i.e., light beams with a non-Gaussian transverse spatial structure and a spatially-varying state of polarization.

\section{Metrological advantage of measuring birefringence with vector vortex beams}
\label{metrological_advantage}
It has been argued that the detection of a small phase unbalance in an interferometric setup can be more precise when measuring the rotation of the petal-like intensity pattern associated to a vector vortex beams than measuring polarization changes~\cite{Emile2017}. The reason provided for this is that the detection of polarization changes usually operates close to total extinction, so the signal power is small. Under the presence of noise, this degrades the signal-to-noise ratio. In comparison, the power of the petal-like intensity pattern structure is not close to zero and is always the same for any phase unbalance, so the signal-to-noise ratio is enhanced and constant. 
 
In general, for a high photon flux and a noise level well above quantum noise (Shot noise), the signal-to-noise ratio is a signature of metrological advantage. However, this is not necessarily the case in close-to-ideal situations where the overall noise level is close to the quantum noise. A fair metrological comparison between measurement schemes requires the use of the appropriate tools of parameter estimation theory. We will make use of the concept of classical Fisher information~\cite{kay1993book,vandeBos2007book}. In a specific measurement scheme, aimed at the estimation of an unknown parameter $\theta$, the number $n$ of photons detected will be associated to a particular probability distribution $p_n(\theta)$, that depends on the parameter $\theta$. The classical Fisher information reads as \cite{kay1993book}
\begin{equation} 
F_c(\theta)=\sum_n p_n(\theta) \left[ \frac{\partial \ln p_n(\theta)}{\partial \theta} \right]^2.
\label{fisher1}
\end{equation}
The important result is that for any estimator $\hat{\theta}$, the precision of the estimation is bounded as (Cramer-Rao bound)
\begin{equation} 
\text{Var}(\hat{\theta}) \geq \frac{1}{F_c(\theta)}
\label{CRbound}
\end{equation}
where $\text{Var}(\hat{\theta})$ is the variance of the statistical estimation. The experimental scheme with higher classical Fisher information can provide, in principle, without considering technological limitations, better precision in phase difference estimation. In appendix~\ref{appendix2}, we demonstrate that when considering coherent states, that is the case in our experiments, the measurement of the mean photon number (power) is an optimum estimator of the phase difference $\theta$, i.e., it attains the Cramer-Rao bound and no other estimator can show better sensitivity. The inequality in Eq.~(\ref{CRbound}) turns out to be an equality when measuring the mean number of photons of a coherent state.
 
\subsection{Sensitivity of the scheme that measures polarization changes of a Gaussian beam}
The number of photons detected in the diagonal $N_D$ and anti-diagonal $N_A$ output ports of $\text{PBS}_2$ (see experimental setup in Fig. \ref{Fig: Montaje}) are
\begin{equation}
N_{D,A}=\frac{N_0 T_0}{2} \left[ 1 \pm \cos \theta \right] \label{photon_number}
\end{equation}
where $N_0$ is the number of photons that illuminate the sample, and $1-T_0$ designates losses. For most molecules with natural activity, if the wavelength of the illumination beam is not resonant with a molecular transition, polarization-dependent losses can be assumed to be negligible. The $\pm$ sign corresponds to $D/A$ polarizations. We assume that the input state is a coherent state. As shown in Appendix~\ref{appendix1}, the quantum state of photons at the output ports of $\text{PBS}_2$ can still be considered coherent states as well, even after experiencing loss. This is a characteristic feature of coherent states, in sharp contrast to most other quantum states of interest.

\begin{figure*}[t!]
  \centering
  \includegraphics[width=\linewidth]{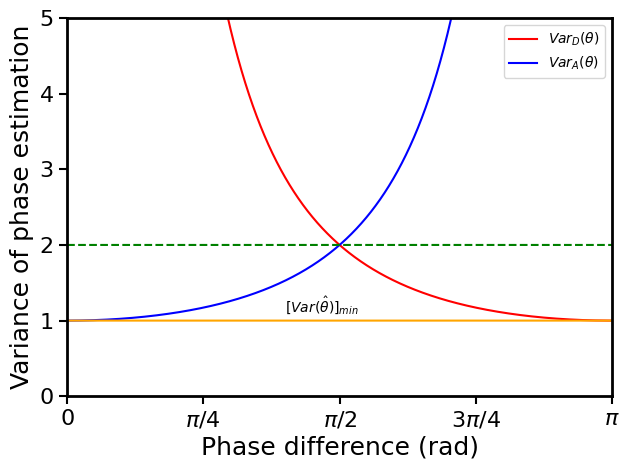}
  \caption{Variance of the statistical estimation of the phase difference as function of the phase difference $\theta$. The values of the variance are normalized to $1/(N _0 T_0)$. Red solid line: projection of the output signal onto diagonal (D) polarization; Blue solid line: projection onto anti-diagonal (A) polarization; Yellow solid line: Minimum Cramer-Rao bound; Dashed green line: measurement of the rotation of a vector vortex beam after projection onto D/A polarizations.}
  \label{Fig:fisher}
\end{figure*}

We need to use Eq. (\ref{fisher1}) to calculate the classical Fisher information associated to the measurement of photons in the diagonal or anti-diagonal ports. Fortunately, for a coherent state, this can be easily done making use of Eq.~(\ref{fisher_coherent}) in Appendix~\ref{appendix2}. It only requires the evaluation of the expression written in Eq.~(\ref{photon_number}), and it avoids the explicit use of the probability distribution. The classical Fisher information reads 
\begin{equation}
F_C^{D,A}=\frac{N_0 T_0}{2}\, \frac{\sin^2 \theta}{1 \pm \cos \theta}
\label{F_A}
\end{equation} 
Since we are considering coherent states, we can write $\text{Var}(\hat{\theta})=1/F_{C}^{D,A}$. Fig.~\ref{Fig:fisher} shows the variance of phase estimation as function of the unknown angle $\theta$ when projecting the output signal onto diagonal (D) and anti-diagonal (A) polarizations. One can easily obtain that the best precision in the estimation of the phase $\theta$ is for the angles that fulfills $\cos \theta=\mp 1$, so $\theta=\pi,0$. Substitution of this value of the angle into Eq. (\ref{F_A}) yields the minimum variance
\begin{equation}
\left[ \text{Var} (\hat{\theta}) \right]_{min}=\frac{1}{N_0 T_0}
\label{F_Amax}
\end{equation}

\subsection{Sensitivity of the scheme that measures the rotation of vector vortex beams}
In Appendix \ref{appendix3} we demonstrate that the classical Fisher information associated to intensity detection with an array of independent pixels is the sum of classical Fisher information for each pixel. In each pixel, located at position ($\rho_i$, $\varphi_i$), we have a coherent state with amplitude $\alpha(\rho_i,\varphi_i)$. The classical Fisher information for pixel $i$, after projection into diagonal/anti-diagonal polarizations, are
\begin{equation}
F_{D,A}^i=\frac{|\alpha(\rho_i)|^2}{2}\,\frac{\sin^2 \left( 2 m \varphi_i+\theta \right)}{1 \pm \cos \left( 2 m \varphi_i+\theta \right)}
\end{equation}
If we approximate discrete cylindrical coordinates by continuous approximations, make use of the integral
\begin{equation}
\int d\varphi\, \frac{\sin^2 (2 m \varphi+\theta)}{1 \pm \cos (2 m \varphi+\theta) }=2\pi
\end{equation}
and take into account that the total number of photons is
$N_0=2\pi\, \int \rho\,d\rho\, |\alpha(\rho)|^2$, we obtain that the classical Fisher information considering all pixels is $F_C^{D,A}=(N_0 T_0)/2$, so the variance of phase estimation is
\begin{equation}
\text{Var} (\hat{\theta})=\frac{2}{N_0 T_0}
\label{F_A2}
\end{equation}
The green dashed line in Fig. \ref{Fig:fisher} depicts the variance for phase estimation using vector vortex beams. If we compare the sensitivity of the two measurement schemes considered here, the first difference is that the precision of phase estimation of the scheme based on vector vortex beams is constant [see Eq. (\ref{F_A2})], it does not depend on the specific value of the unknown parameter $\theta$. It also does not depend on projecting onto diagonal or anti-diagonal polarization. On the contrary, the precision of the scheme that makes use of a Gaussian beam is $\theta$-dependent, as shown in Eq. (\ref{F_A}). In this case, there is a phase $\theta$ where the precision is minimum (classical Fisher information is maximum), that depends on the polarization that we detect.

When we project the output signal onto diagonal polarization, the scheme that makes use of vector vortex beams achieve a better precision for angles $\theta \leq \pi/2$. If we project onto anti-diagonal polarization, this is the case for $\theta \geq \pi/2$. When the {\it standard} scheme, which measures polarization changes, works at the corresponding optimum angle ($0$ or $\pi$), it achieves a factor of $2$ metrological advantage in phase estimation. 

If we consider the measurement of the signal in both output ports of $\text{PBS}_2$ [see Fig. \ref{Fig: Montaje}], the associated classical Fisher information is the sum of classical Fisher information for projections onto D/A polarizations, since we are considering coherent states. For both experimental schemes considered here, we obtain the same result $F_C=F_C^D+F_C^A=N_0 T_0$. We can say that from a fundamental perspective, without considering the technological limitations relevant for each scheme, birefringence metrology measuring polarization changes, or rotations of the petal-like intensity pattern, can provide the same sensitivity in phase estimation.

\section{Experimental setup}
\label{experimental_setup}
\begin{figure*}[t!]
  \centering
  \includegraphics[width=1\linewidth]{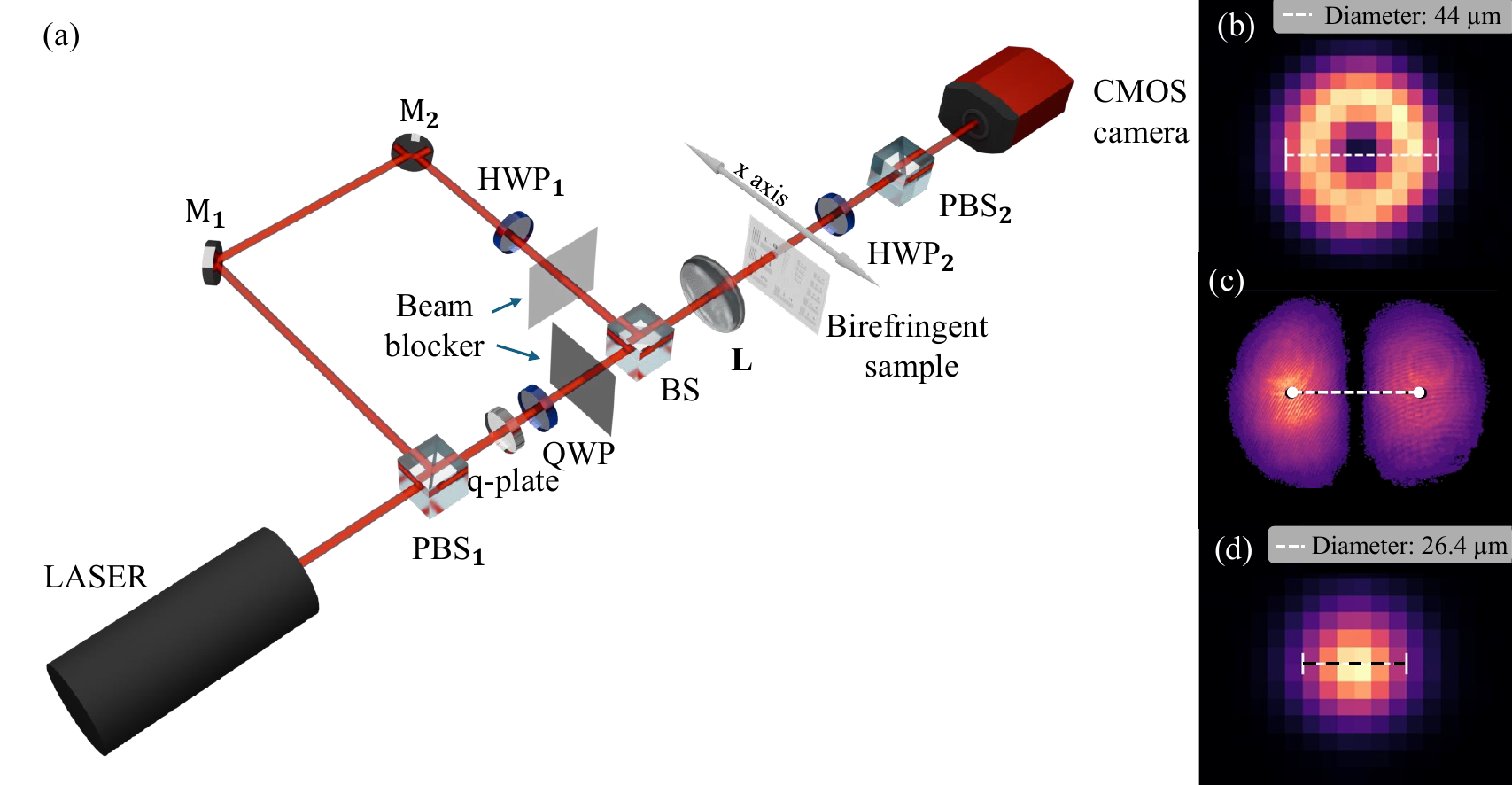}
  \caption{(a) Experimental setup to measure birefringence with two different schemes. In order to choose each measurement scheme, the setup consists of two optical paths that can be blocked independently. In one path, the illumination beam is a diagonally-polarized Gaussian beam. In the other path, a q-plate generates an optical beam where the spatial shape associated to each orthogonal polarization component (horizontal/vertical) carries OAM with topological charge $m=\pm1$. (b) Intensity profiles at the sample plane of the optical beam carrying OAM. The diameter of the doughnut beam is $44.0~\mu \text{m}$. (c) Intensity profile of the optical beam carrying OAM at the detection plane, after projection onto diagonal polarization with the help of $\text{HWP}_2$ and $\text{PBS}_2$. (d) Intensity profile of the optical beam with a Gaussian shape spatial profile at the sample plane. The beam waist of the optical beam is $26.4~\mu \text{m}$. }
  \label{Fig: Montaje}
\end{figure*}

Figure~\ref{Fig: Montaje}(a) shows the experimental setup used to measure birefringence. For the sake of comparison, the setup is designed to measure birefringence using two different measurement schemes: the standard scheme, that detects polarization changes of a Gaussian beam, and the alternative scheme we put forward here, that makes use of a vector vortex beam. A HeNe laser operating at $633~\text{nm}$ generates a diagonally-polarized optical beam with a Gaussian spatial shape. The optical beam is directed onto a Polarizing Beam Splitter ($\text{PBS}_1$) to generate two orthogonally-polarized beams that propagate along different paths. The beam transmitted at $\text{PBS}_1$ traverses a q-plate, that in combination with a quarter-wave plate (QWP), generates a vector vortex optical beam, that writes
\begin{equation}
\mathbf{E}_{\text{out}}(\rho,\phi) = \frac{1}{\sqrt{2}}\,E_0(\rho) \left[ \exp (im\varphi)\,\hat{H} + \exp(-im\varphi\, \hat{V} )\right] 
\end{equation}
whose $(\rho,\varphi$) are cylindrical coordinates, $\hat{H}\,$/$\,\hat{V}$ designate horizontal/vertical polarizations, and $E_0(\rho)$ is the radial profile of the optical beam leaving the q-plate. In our experiments, we make use of a q-plate with $q=1/2$, that generates a vector vortex beam where each orthogonal polarization component is a vortex beam with topological charge $m=\pm 1$. After the beam splitter (BS), a lens with focal length $L=50~\text{mm}$ focuses the optical beam onto the sample plane. Fig.~\ref{Fig: Montaje}(b) depicts the spatial intensity distribution of the vector vortex beam measured by locating the CMOS camera in the sample plane. The doughnut shape has a diameter of $44~\mu \text{m}$, measured as the distance between the outer pixels with a $1/e^2$ drop in intensity with respect to the spatial locations with the peak intensity.

Light traverses a birefringent sample [\textit{NBS 1963A Birefringent Resolution Target} (Thorlabs)] that introduces a phase difference $\theta(x)$ between horizontal and vertical polarizations. The sample can be scanned in the $x$ direction. Finally, with the help of half-wave plate $\text{HWP}_2$ and polarizing-beam splitter $\text{PBS}_2$, the output vector vortex beam is projected into diagonal polarization ($\text{HWP}_2$ at $22.5^\circ$) or anti-diagonal polarization (at $-22.5^\circ$). The spatial intensity distribution measured using the CMOS camera, after projection onto diagonal polarization, is
\begin{equation}
I = I_0(\rho) \cos^2 \left( m\varphi + \frac{\theta}{2} \right),
\label{oam_light}
\end{equation}
where $I_0(\rho)=|E_(\rho)|^2$. Fig.~\ref{Fig: Montaje}(c) shows an example of the spatial intensity distribution at the detection plane. The output light beam covers almost all of the area of the CMOS camera ($12.6~\text{mm} \times 12.6~\text{mm}$) due to the divergence induced by lens $L$. In general, the optical beam at the detection plane consists of $2m$ lobes, that we will refer as {\it petals}. For $m=\pm 1$, the case we consider here, we can observe two petals. The unknown phase difference $\theta$ causes a rotation of the intensity pattern distribution of $\Delta \varphi=\theta/(2m)$. The total power after projection onto diagonal (anti-diagonal) polarizations is $P_0/2$, where $P_0=\int \rho\, d\rho\, I_0(\rho)$ is the power of the input beam. Notice that the power after projection is independent of the specific value of $\theta$.

\begin{figure*}[t!]
  \centering
  \includegraphics[width=0.9\linewidth]{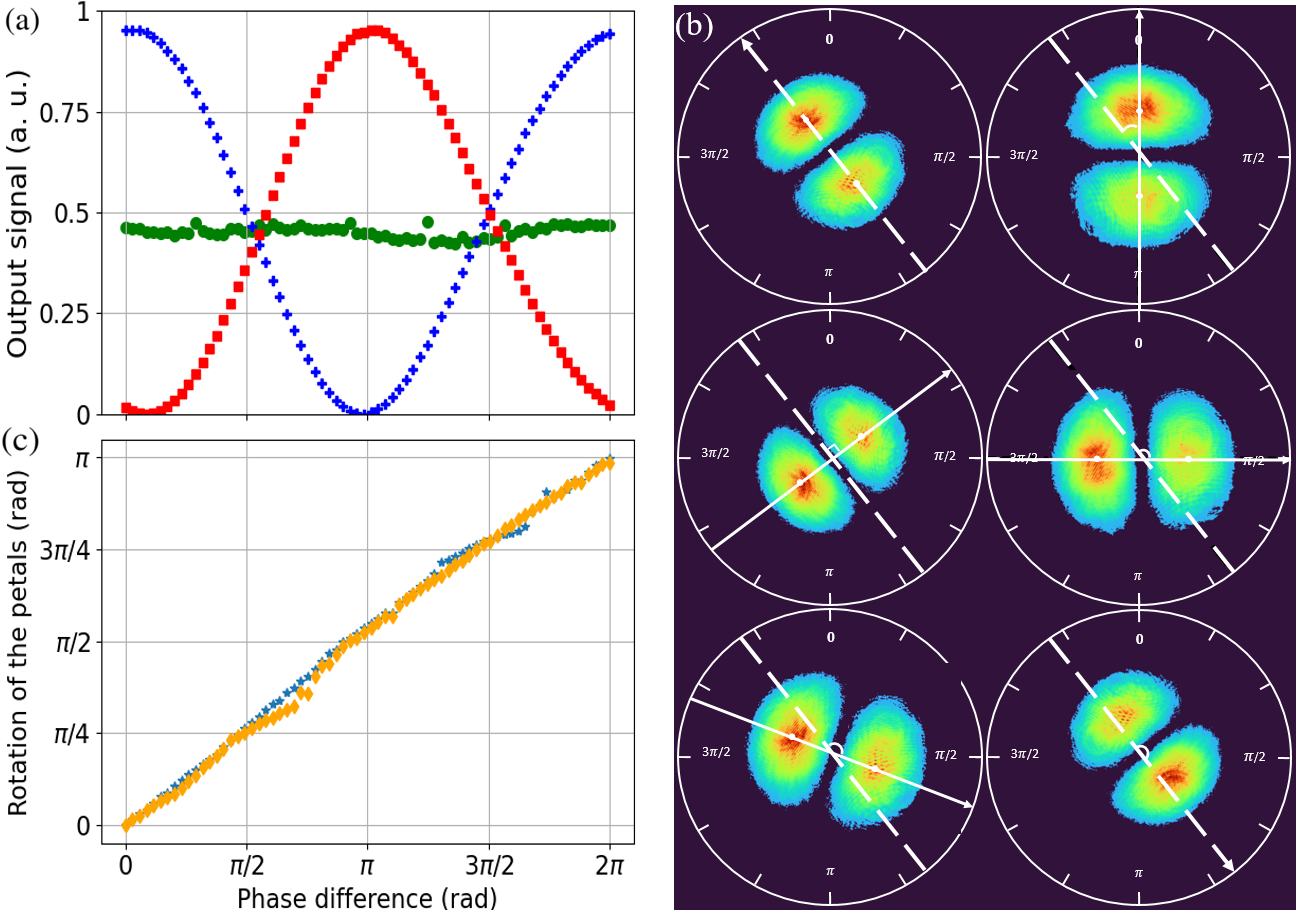}
  \caption{(a) Output power $P$ measured as function of the phase difference $\theta$ introduced by the Liquid Crystal Variable Retarder. Red and blue squares: Gaussian beam after projection onto diagonal (blue) and anti-diagonal (red) polarization. Green dots: vector vortex beam after projection onto diagonal polarization. (b) Six examples of the rotation of the intensity pattern of the output optical beam at the detection plane, for different values of $\theta$. (c) Value of the rotation of the petals as function of $\theta$. respectively.}
  \label{Fig:Caracterization}
\end{figure*}
The polarization of the Gaussian beam reflected at $\text{PBS}_1$ is set to diagonal by a half-wave plate $\text{HWP}_1$ set at $22.5^\circ$. After being reflected at the beam splitter, lens $L$ focuses the Gaussian beam onto the sample plane. Fig.~\ref{Fig: Montaje}
(d) shows the spatial distribution of the Gaussian mode in the sample plane, measured by placing the CMOS camera in the sample plane. The Gaussian beam has a diameter of $26.4\,\mu\text{m}$ ($1/e^2$ beam width). The power $P$ of the output optical beam, after projection into diagonal polarization, is 
\begin{equation}
  P=P_0 \cos^2 \frac{\theta}{2},
  \label{gaussian_light}
\end{equation}
The output power follows a $sin$ function instead, after projection onto anti-diagonal polarization.

Eqs.~(\ref{oam_light}) and (\ref{gaussian_light}) will be used to estimate the unknown phase difference $\theta$. Beam blockers located in the transmitted and reflected beams paths after $\text{PBS}_1$ are used to switch between the two measurement schemes.

\section{Experimental results}
\label{experimental_results}
In order to characterize and calibrate the two measurement schemes, we first locate in the sample plane a Liquid Crystal Variable Retarder (LCVR), which introduces a controllable phase difference $\theta$ between the polarizations $\hat{H}\,$/$\,\hat{V}$ by varying the voltage applied to the LCVR. Fig.~\ref{Fig:Caracterization} shows the experimental results obtained using the two measurement schemes. A particular scheme is selected by blocking the path corresponding to the alternative measurement. 

\begin{figure*}[t!]
  \centering
  \includegraphics[width=0.9\textwidth]{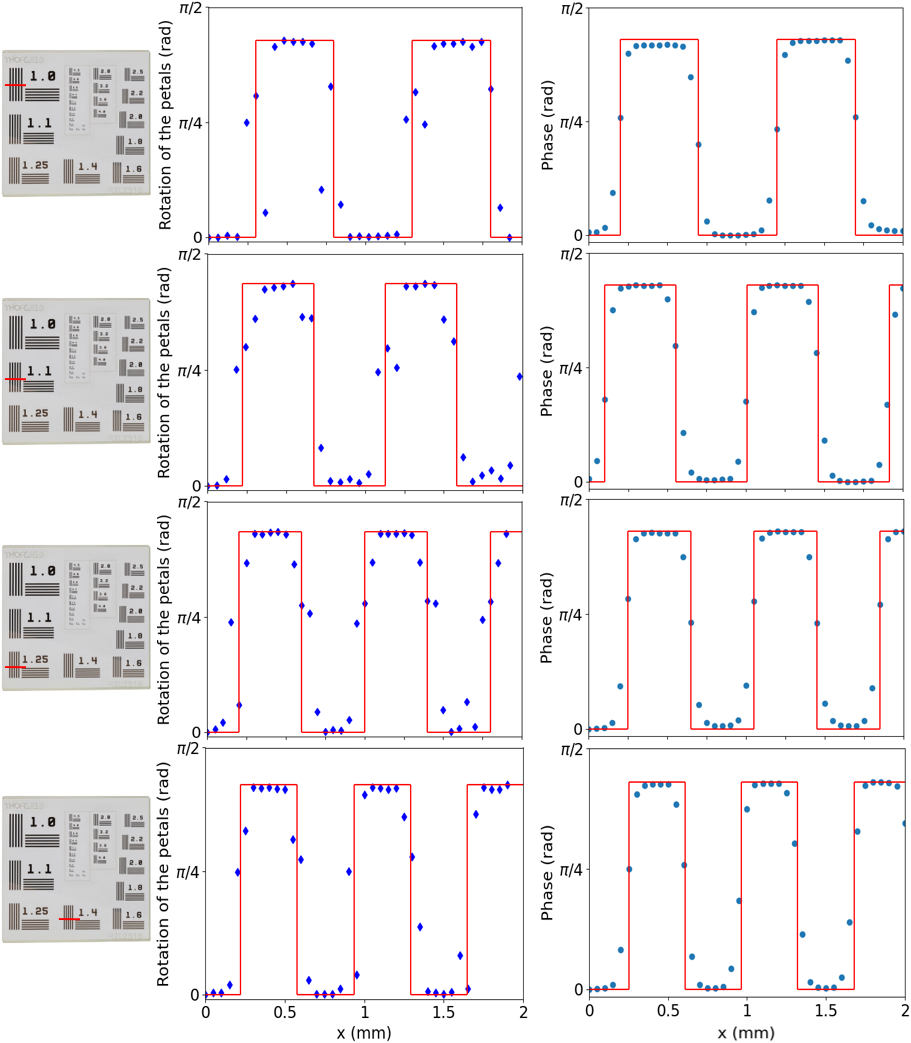}
  \caption{Measurement of birefringence using the NBS 1963A Birefringent Resolution Target (Thorlabs). The four images shown on the left depicts the birefringent target used in the experiment, indicating with a red solid line the specific region scanned (transverse position $x$). Measurements shown in the left column corresponds to the case when we estimate the phase difference $\theta$ measuring the rotation of the intensity pattern generated with the help of a vector vortex beam. Measurements shown in the right column corresponds to the standard case when we estimate the phase difference measuring the power of the output power of a Gaussian beam. The blue markers show the experimental data. The red solid lines are best fits considering a periodic rectangular step function, whose free parameters are the amplitude and the spatial frequency. The values of the best fit for the experimental data shown on the left column are $ \text{height (rad)} = \left[ 1.34, 1.37, 1.36, 1.34 \right]$ and $ \text{frequency (cycles/mm)} = \left[ 1.01, 1.10, 1.24, 1.39 \right]$. The values of the best fit for the experimental data shown on the right column are $\text{(height (rad)} = \left[ 1.32, 1.29, 1.28, 1.35 \right]$ and $\text{frequency (cycles/mm)} = \left[ 1.01, 1.12, 1.25, 1.42 \right]$.}
  \label{Fig:Sample}
\end{figure*}

The experimental curves represented by red squares and blue crosses in Fig.~\ref{Fig:Caracterization}(a) show the power of the output Gaussian beam measured as a function of the phase difference $\theta$, after projection onto diagonal (blue) and anti-diagonal polarizations (red). These measurements were done replacing the CMOS camera [see Fig.~\ref{Fig: Montaje}(a)] with a photodiode and storing the values measured with a FPGA board for each value of the phase. The experimental results exhibit the oscillatory behavior as expected from Eq.~\ref{gaussian_light}. The curve represented by green dots corresponds to the power measured using the vector vortex beam, after projection onto diagonal polarization. These measurements were done by taking a picture with the CMOS camera for each value of the phase difference, and integrating the intensity measured in all the pixels of the image. The power remains constant regardless of the value of the phase difference $\theta$, as expected from integration of Eq.(\ref{oam_light}). 

The phase difference introduced by the LCVR causes a rotation of the petals that form the intensity pattern of the beam resulting after projection of the vector vortex beam onto diagonal/anti-diagonal polarization. Fig.~\ref{Fig:Caracterization}(b) shows six examples of this rotation for different values of the phase difference $\theta$ introduced. For each image, a computational algorithm was employed to determine the center of mass of each petal~\cite{ye2024,cao2024}. A solid line joins the centers of mass of each petal of the beam, which is perpendicular to the line of zero intensity that corresponds to $\Delta \varphi=\theta/2$. The arrow of the solid line indicates the angle of the center of mass. The upper-left image serves as a reference. The dashed line indicates the initial orientation and is retained in all subsequent images to facilitate visualization of the rotation observed in each case. In the remaining images, the solid line indicates the orientation of the petal pattern after applying a voltage to the LCVR.

Figure~\ref{Fig:Caracterization}(c) depicts the relative rotation of the petals as function of the phase difference $\theta$ introduced by the LCVR. Each point in the $y$-axis is obtained by subtracting the rotation observed when the voltage applied is zero, from the rotation observed for each value of the voltage. Orange/blue points correspond to the projection onto diagonal/anti-diagonal polarizations, respectively. The straight lines are linear fits that agree with the expected slope~$\approx 1/2$.

In order to demonstrate the feasibility of the scheme introduced here, based on the use of vector vortex beams, on a {\it real} birefringent sample, we make use of the {\it NBS 1963A Birefringent Resolution Target} from Thorlabs. It consists of a birefringent pattern sandwiched between a glass substrate and a protective glass, both made from N-BK7. 

The images on the left of Fig.~\ref{Fig:Sample} correspond to the sample. The red lines indicate the scan path. The number shown beside the line is the spatial frequency in $cycles/mm$ of the dark and bright sections of the target. We performed experiments with patterns with spatial frequencies from $1$ to $1.4$ cycles/mm, which corresponds to a period of $1~\text{mm}$ to $714~\mu\text{m}$. The column with experimental data on the left are the results using the vector vortex beam. The scan step size is $\Delta x=0.5~\text{mm}$. During the scan, the bright and dark bars of the sample produce a rotation of the petal-like intensity pattern structure of around $\Delta \varphi=1.34~\text{rad}$. This corresponds to a measurement of the phase difference of $\theta=0.67~\text{rad}$, since $\Delta \varphi=\theta/2$ for a vortex beam with $m=1$. The solid line is a fit using the spatial frequency of each segment as a fixed parameter. The third column is the scan measuring the polarization of a Gaussian beam. In this case, the change in intensity of the output beam indicates the value of the phase difference. We make use of Fig.~\ref{Fig:Caracterization}(a) to identify the relationship between intensity and phase.

Fig.~\ref{Fig:Caracterization}(a) highlights an important difference between the two measurement schemes compared here. In the standard scheme, for each output port (projection onto diagonal or anti-diagonal polarization), the output power depends on the unknown phase difference $\theta$. When using a vector vortex beam, the output power remains constant for all values of $\theta$. This difference has been cited as a fundamental advantage of the scheme using a vector vortex beam, since it allows to detect a higher number of photons, reducing the uncertainty~\cite{Emile2017}.

\section{Conclusions}
\label{conlusions}
We have demonstrated experimentally that one can estimate accurately, in a reliable way, the amount of birefringence introduced by a sample, as it is the case in Differential Interference Contrast microscopy or Chiral analysis, using vector vortex beams. We have compared the results obtained from vector vortex beams with results obtained using the conventional procedure of measuring the change of the global state of polarization of a Gaussian beam. 

An advantage of the technique put forward here is that the precision of phase estimation is independent of the particular value of the unknown phase $\theta$ that we want to estimate. We have also demonstrated that the measurement of the rotation of the petal-like intensity pattern structure of a light beam can provide a sensitivity enhancement (better precision in phase estimation). In this case, the fundamental limitation comes from our ability to measure accurately tiny rotations of the intensity pattern with current CMOS cameras.

Structured light has emerged in the last few decades as an important resource in modern optics. It has attracted significant attention in both classical and quantum optics~\cite{torner20011book,halina2013,Bliokh2023}. In particular, there is a myriad of applications of structured light for imaging and sensing~\cite{tk2004,gozali2017,Sokolenko2018,zhao2021,cheng2025}. The spatial shape of light beams provides a multidimensional alphabet~\cite{wang2012} where its complex spectrum can be used as a resource for imaging~\cite{Torner2005,Hermosa2014,Xie2017}. The experiments and theoretical calculations presented here can be considered as an example where the use of non-separable light beams, where the polarization and spatial degrees of freedom can not be considered independently, provides a metrological advantage for sensing and estimation, not attainable using conventional procedures based on the use of separable light beams. 

\appendix

\section{Evaluation of the classical Fisher information associated to a coherent state}
\label{appendix2}
The probability distribution of a single-mode coherent state corresponds to a Poisson distribution
\begin{equation}
p_n(\theta)=\exp \left[ -\lambda(\theta) \right] \frac{ \left[ \lambda(\theta) \right]^n}{n!}
\label{poisson}
\end{equation}
where $n$ is the number of photons, $\lambda(\theta)$ is the mean number of photons, and $\theta$ is the parameter whose value we want to estimate. The classical Fisher information is
\begin{equation} 
F_c=\sum_n p_n \left( \frac{\partial \ln p_n}{\partial \theta}\right)^2.
\end{equation}
If we use Eq. (\ref{poisson}), we have $\ln p_n=-\lambda(\theta) + n \ln \lambda(\theta)-\ln n!$. Taking into account that for the Poisson distribution, $\sum_n n p_n=\lambda$ and $\sum_n n^2 p_n=\lambda^2+\lambda$, it is straightforward to show that the classical Fisher information can be simply evaluated as 
\begin{equation}
F^C=\frac{1}{\lambda(\theta)} \left[ \frac{\partial \lambda(\theta)}{\partial \theta} \right]^2
\label{fisher_coherent}
\end{equation}
This is the same expression that one would obtain using the propagation of errors equation. Therefore, the Cramer-Rao bound can be attained by measuring the mean number of photons. Measurement of the power is an optimum estimator of the unknown parameter $\theta$, no other estimator can surpass its sensitivity. 

\section{Nature of a coherent state that experiences loss}
\label{appendix1}
The main effect of losses on a quantum state can be mimicked~\cite{boyd2008} making use of a beam splitter with transmissivity $\sqrt{T}$ and reflectivity $\sqrt{R}$, with $R+T=1$. The input ports are $1$ and $2$, the output ports are $3$ and $4$. Output port $4$ is the loss reservoir. The amount of loss is $R$. If there is a single-mode coherent state with amplitude $\alpha$ in input port $1$, the quantum input state can be written as $|\Psi \rangle{12}=|\alpha \rangle_1 | 0\rangle_2=D_1(\alpha)|0 \rangle_1|0\rangle_2$, where
\begin{equation}
D_1(\alpha)=\exp \left( \alpha a_1^{\dagger}-\alpha^* a_1 \right)
\end{equation}
$a_1$ and $a_1^\dagger$ are the annihilation/creation operators
associated to input port $1$, and $D$ is the displacement operator. The input-output relationship between input and output operators in the beam splitter are
\begin{eqnarray}
& & a_1 \Longrightarrow \sqrt{T} a_3+i\sqrt{R} a_4 \nonumber \\
& & a_2 \Longrightarrow i\sqrt{R} a_3+\sqrt{T} a_4
\end{eqnarray}
The quantum state at the output ports of the beam splitter is
\begin{widetext}
\begin{equation}
|\Psi \rangle_{34}=\exp \left\{ \alpha \left[ \sqrt{T} a_3^{\dagger}-i \sqrt{R} a_4^{\dagger} \right] \alpha^* \left[ \sqrt{T} a_3+i \sqrt{R} a_4 \right] \right\}\, |0 \rangle_3|0\rangle_4
\end{equation}
\end{widetext}
Since the operators $a_3$ and $a_4$ commute, we can write

\begin{align}
|\Psi \rangle_{34}&=\exp \left\{ \alpha \sqrt{T} a_3^{\dagger}- \alpha^* \sqrt{T} a_3 \right\}\, \\ \nonumber  &\hspace{0.7cm} \times \exp \left\{ -i \sqrt{R} a_4^{\dagger} - i \alpha^* \sqrt{R} a_4 \right\}\,|0 \rangle_3|0\rangle_4 \\ \nonumber 
 &=|\alpha \sqrt{T}\rangle_3\, |-i \alpha \sqrt{R}\rangle_4 
\end{align}

The main result is that losses do not change the quantum nature of the coherent state, only its amplitude. A coherent state remains a coherent state even under the presence of loss. This is an important result that we will use below when considering the effect of loss for the two measurement schemes considered here. 

\section{Classical Fisher information for a multimode coherent state}
\label{appendix3}
Let us consider an array of $M$ pixels. The probability distributions associated to the number of photons $n_i$ arriving to each pixel ($i=1 \dots M$) are assumed to be independent, so the overall probability $p_{\alpha}$ associated to a particular detection event $\alpha=n_1 \dots n_i \dots n_M$ photons in each pixel is
\begin{equation}
p_{\alpha}=p_1(n_1) \dots p_i(n_i) \dots p_M(n_M)
\end{equation}
The classical Fisher information is
\begin{equation}
F_c=\sum_{\alpha} p_{\alpha} \left( \frac{\partial \ln p_{\alpha}}{\partial \theta}\right)^2.
\end{equation}
where $\ln p_{\alpha}=\sum_i \ln p_i$. We can write

\begin{align}
F_c&=\sum_i\, \sum_{\alpha} p_1 \dots p_i \left( \frac{\partial \ln p_i}{\partial \theta} \right)^2 \dots p_M \\ \nonumber
&+2\sum_{i \ne j}\, \sum_{\alpha} p_1\dots p_i \frac{\partial \ln p_i}{\partial \theta}\dots p_j \frac{\partial \ln p_j}{\partial \theta}\dots p_M 
\end{align}

Making use of $\sum_{n_i} p_i=1$ and $\sum_{n_i} p_i \left( \partial \ln p_i/\partial \theta \right)=0$, we obtain
\begin{equation}
F_c=\sum_i\, p_i \left( \frac{\partial \ln p_i}{\partial \theta} \right)^2=\sum_i F_c^i 
\end{equation}
The classical Fisher information associated to the array of $M$ pixels is the sum of the classical Fisher information associated to each pixel.

\bibliography{References}


\section*{Funding} This work was partially funded by CEX2024-001490-S [MICIU/AEI/10.13039/501100011033]. CIO; SECIHTI. G.F.C. (CVU: 1316339) and D.S.R. (CVU: 1345671) acknowledge funding from Secretaría de Ciencia, Humanidades, Tecnología e Innovación (SECIHTI) granted through their PhD’s scholarships, with identification codes INV-2022-142-2435 and INV-2022-143-2490. 

\section*{Acknowledgment}
We acknowledge support from the project NOVISLIGHT (PID2023-149780NB-I00) funded by Ministerio de Ciencia, Innovación y Universidades (Proyectos de generación de conocimiento 2023). This work is part of the R$\&$D project CEX2019-000910-S, funded by the Ministry of Science and innovation (MCIN/AEI/10.13039/501100011033/). It has also been supported by Fundació Cellex, Fundació Mir-Puig, and from Generalitat de Catalunya through the CERCA program. 




\end{document}